\documentclass[referee]{raa}

\usepackage{graphicx,times}
\usepackage{natbib}
\usepackage{amssymb,amsmath}
\bibpunct{(}{)}{;}{a}{}{,}
\usepackage{tikz}
\usetikzlibrary{shapes.geometric, arrows, positioning}
\usepackage[colorlinks=true,linkcolor=blue,citecolor=blue,pagebackref=true]{hyperref}
\usepackage{orcidlink}
\usepackage{longtable} 
\usepackage{multirow}   
\newcommand{\hb}{H$\beta$}

\begin{document}

   \title{Accretion-Mode Transition: The Driver Behind Spectral Changes in Changing-Look AGNs}

   \volnopage{Vol.0 (20xx) No.0, 000--000}
   \setcounter{page}{1}

   \author{
      Pengyuan Han\inst{1}
   \and 
      Huaiyuan Lu\inst{1}
   \and 
      Bing Lyu$^{*}$\inst{2} \orcidlink{0000-0001-8879-368X}
   \and 
    Jiancheng Wu\inst{3} \orcidlink{0000-0002-2581-8154} 
   \and    
      Qingwen Wu$^{*}$\inst{1} \orcidlink{0000-0003-4773-4987}
   }
   
   \institute{
   		\inst{1} Department of Astronomy, School of Physics, Huazhong University of Science and Technology, Wuhan 430074, China; {\it qwwu@hust.edu.cn} \\
      \inst{2} Kavli Institute for Astronomy and Astrophysics, Peking University, Beijing 100871, China; {\it lyubing@pku.edu.cn} \\ 
      \inst{3} Institute for Astronomy, School of Physics, Zhejiang University, 866 Yuhangtang Road, Hangzhou 310058, China \\ 
      \vs\no
      {\small Received 20xx month day; accepted 20xx month day}
   }

\abstract{
The physical origin of optical changing-look AGNs (CLAGNs), characterized by the appearance or disappearance of broad emission lines, is thought to be mainly driven by the variation of the black-hole (BH) accretion rate. In this work, we explore this issue based on a sample of {224} CLAGNs with UV-to-optical continua, where the UV radiation is more sensitive to the accretion state near the BH horizon.  We find that the luminosity correlation of $L_{3000}$--$L_{5100}$ at 3000$\rm \AA$ and 5100$\rm \AA$ becomes steeper at low luminosities (e.g., $L_{3000}\lesssim10^{44}\rm erg/s$), where the sources with high luminosities are roughly consistent with the prediction of a standard accretion disk. At lower luminosities, the observations are more consistent with the prediction of a truncated disk. The whole sample has a median bolometric Eddington ratio of $\sim$2.2\%, which is consistent with the critical value for state transition in X-ray binaries. Such transitions can significantly alter the UV-to-optical continuum, largely due to variations in the truncation radius, even when the change in the overall accretion rate is minimal. The deficit of ionization photons resulting from an increase in the truncation radius will lead to the weakening or disappearance of broad lines, which triggers the AGN changing-look. 
\keywords{accretion, accretion disks --- galaxies: active --- quasars: general --- quasars: supermassive black holes}
}

   \authorrunning{P.-Y. Han et al.}
   \titlerunning{Accretion-Mode Transition in CLAGNs}

   \maketitle

\section{Introduction}
\label{sect:intro}

Active galactic nuclei (AGNs) are powered by the accretion of matter onto supermassive black holes (SMBHs) at the centers of galaxies \citep{1969Natur.223..690L}. The standard paradigm, known as the AGN unified model, successfully explains the observed dichotomy between type 1 AGNs (showing both broad and narrow emission lines) and type 2 AGNs (showing only narrow lines) as a consequence of different viewing angles \citep[e.g.,][]{1993ARA&A..31..473A, 1995PASP..107..803U}. In the AGN unified model, a dusty torus obscures the broad line region (BLR) in type 2 AGNs with high-inclination lines of sight, while the BLR is visible in type 1 AGNs with low-inclination lines of sight. However, this orientation-based unified model has been challenged by the growing number of changing look AGNs (CLAGNs). These objects exhibit transitions between type 1 and type 2 classifications on timescales of months to years . The advent of large spectroscopic surveys has enabled systematic searches for these events \citep[e.g.,][]{2016MNRAS.457..389M,2018ApJ...862..109Y,2023MNRAS.518.2938T,2024ApJ...966...85Z,2025ApJ...980...91Y,2025ApJ...986..160D,2025ApJS..278...28G}, moving beyond serendipitous discoveries \citep[e.g.,][]{2015ApJ...800..144L,2020MNRAS.491.4925G}, and even employing novel methods based on identifying mismatches between historical spectral types and modern photometric variability patterns \citep[e.g.,][]{2024ApJ...966..128W,2025MNRAS.536.2715Z}.

Several physical mechanisms have been proposed to explain the changing-look phenomenon \citep[see a recent review in][]{2023NatAs...7.1282R}. While a small portion of changing-look events may be triggered by external perturbations like tidal disruption events (TDEs; e.g., \citealt{2021ApJS..255....7R,2024ApJ...975...50L}), or change of obscuration \citep[e.g.,][]{2007MNRAS.377..607P,2025MNRAS.537.1099L}, the leading explanation for the majority of CLAGNs is due to an intrinsic change in the accretion rate \citep[e.g.,][]{2017ApJ...846L...7S,2018MNRAS.480.3898N,2021MNRAS.506.4188L,2022JHEAp..33...20L,2024ApJ...970...85W,2025ApJS..276...51L}. In the high state, the accretion disk is believed to be a geometrically thin, optically thick standard disk \citep[SSD;][]{1973A&A....24..337S}, which is a strong emitter of optical and UV photons that ionize the BLR. In the low state, the inner region of accretion flow might transit into a radiatively inefficient state, such as an Advection-Dominated Accretion Flow (ADAF; \citealt{1994ApJ...428L..13N}). The deficit of UV-to-X-ray ionizing photons is unable to excite the gas in the BLR. The variations and delays of mid-infrared emission in CLAGNs 
indicate that the central ionization luminosity of CLAGNs is changed \citep{2022ApJ...927..227L}. The growing number of repeating or recurrent CLAGNs, which experience multiple changing look events, supports the strong variation of UV ionization spectrum in them \citep[e.g.,][]{2025A&A...693A.173L, 2025arXiv251018445D, 2025ApJ...981..129W, 2025A&A...698A.135G}.

The UV continuum emitted from the innermost regions of the accretion disk is one of the most effective tracers of structural changes in the accretion flow and the primary energy source for ionizing the BLR. Therefore, analyzing the evolution of the UV-to-optical spectral shape during state transitions provides a powerful diagnostic tool for changing look phenomena. In this paper, we compile a sample of CLAGNs with multi-epoch spectroscopy from the Sloan Digital Sky Survey (SDSS; \citealt{2022ApJS..259...35A}), the Dark Energy Spectroscopic Instrument (DESI; \citealt{2024AJ....168...58D}), and the Large Sky Area Multi-Object Fiber Spectroscopic Telescope (LAMOST; \citealt{2015RAA....15.1095L}) to investigate their UV-optical continuum. We compare our observational results with the predictions from both the standard disk and the truncated disk models to study the physical driver of the changing-look phenomenon. The outline of this paper is as follows. In \autoref{sect:data_analysis}, we present our sample and spectroscopic fitting. The results are presented in \autoref{sect:results}. We discuss our results in  \autoref{sect:discussion} and summarize in  \autoref{sect:conclusion}. 

\section{Sample and Data Analysis}
\label{sect:data_analysis}

\subsection{Sample Compilation and Spectroscopic Data}
The parent sample for this study is constructed by compiling a comprehensive list of CLAGNs previously identified in the literature. Our initial candidate list is primarily drawn from recent large, systematically-searched catalogs, building upon foundational works such as \citet{2016MNRAS.457..389M}. Specifically, our sample combines sources from three major recent surveys:
\begin{itemize}
    \item \textbf{130} sources from the DESI early data sample of \citet{2024ApJS..270...26G}, which represents one of the first large-scale CLAGN searches with DESI.
    \item \textbf{592} sources from the DESI DR1 sample presented in the successor paper by \citet{2025ApJS..278...28G}, which significantly expands the number of DESI-identified CLAGNs.
    \item \textbf{51} sources from the SDSS/LAMOST sample of \citet{2025ApJ...986..160D}, providing a complementary set of objects with long time baselines.
\end{itemize}

For each source in this compiled list, we performed a systematic search for all available spectroscopic observations from the three major public archives (SDSS, DESI, and LAMOST). This cross-matching process yielded an initial collection of over a thousand spectroscopic epochs. 

\subsection{Spectral Fitting Methodology}
\begin{figure}
   \centering
   \includegraphics[width=1.0\textwidth]{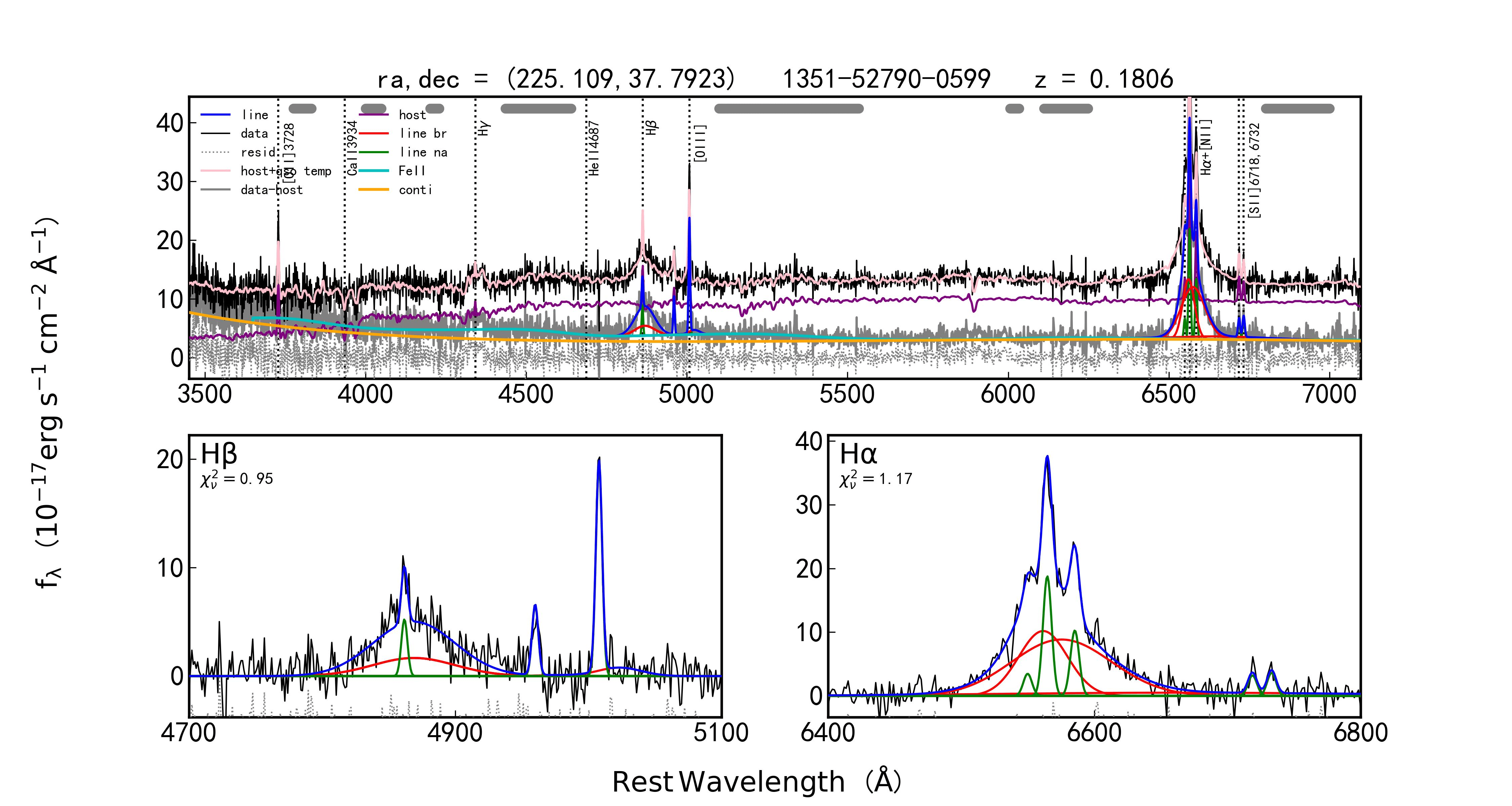}
   \caption{An example of the spectral fitting for a CLAGN spectrum using \texttt{PyQSOFit}. The observed spectrum is shown in black. The best-fit model is in red, composed of a power-law continuum (blue), Fe~II templates (green), and Gaussian emission line components (orange). }
   \label{fig:fit_example}
\end{figure}

All spectra in our collection were re-fitted using the public Python code \texttt{PyQSOFit} \citep{2018ascl.soft09008G, 2019ApJS..241...34S}, which was also successfully applied to LAMOST quasar samples \citep[e.g.,][]{2023ApJS..265...25J}. The spectra were corrected for Galactic extinction using the dust maps of \citet{1998ApJ...500..525S} and assuming a Milky Way extinction law with $R_V = 3.1$ \citep{1989ApJ...345..245C}. The fitting process involves decomposing each spectrum into a power-law continuum, a host galaxy component, empirical Fe~II templates \citep{2001ApJS..134....1V, 1992ApJS...80..109B}, and a series of Gaussian functions for the emission lines. An example of the spectral decomposition is shown in Figure~\ref{fig:fit_example}.

From the best-fit models, we measured the continuum luminosities and the properties (FWHM and luminosity) of the broad emission lines. The single-epoch virial black hole masses ($M_{\rm BH}$) are then estimated using the scaling relations based on the H$\beta$ line \citep{2006ApJ...641..689V}. We simply estimated the bolometric luminosities ($L_{\rm bol}$) from 5100\,\AA\ luminosity using the empirical correlation of $L_{\rm bol}\sim 10 L_{5100}$ \citep{2019ApJ...886...42D}. The bolometric Eddington ratio can be derived as $\lambda_{\rm Edd} = L_{\rm bol} / L_{\rm Edd}$, where $L_{\rm Edd} = 1.26 \times 10^{38} (M_{\rm BH}/M_\odot)$ erg s$^{-1}$. 


We then applied a series of rigorous quality cuts to ensure the reliability of our measurements. The step-by-step procedure of this data reduction is summarized in Table~\ref{tab:sample_reduction}. The main criteria include a redshift cut to focus on our primary analysis range, basic quality cuts on signal-to-noise and host galaxy contamination, and a requirement for a good quality spectral fit. This process ensures that any observed variations are significant and not dominated by measurement uncertainties. After these steps, we selected the final sample for our primary science goal, which required sources to have reliable measurements of both $L_{3000}$ and $L_{5100}$. Our final sample used for analysis consists of 276 high-quality spectra from 224 unique CLAGNs. The properties of this sample are listed in Table~\ref{tab:sample_properties}, where the first three entries are shown for illustration.

\begin{table}[h!]
\centering
\caption{Summary of the Data Reduction and Sample Selection Steps.}
\label{tab:sample_reduction}
\setlength{\tabcolsep}{5pt} 
\begin{tabular}{clcc}
\hline\hline
\noalign{\smallskip}
Step & Criterion & No. of Spectra & No. of Sources \\
\noalign{\smallskip}
\hline
\noalign{\smallskip}
1 & Initial parent sample from literature & 1,058 & 592 \\
  & \textit{\scriptsize \hspace{1em} $\rightarrow$ Apply redshift cut ($z < 1.0$)} & & \\
2 & After redshift cut & 1,034 & 568 \\
  & \textit{\scriptsize \hspace{1em} $\rightarrow$ Apply basic quality cuts\textsuperscript{a}} & & \\
3 & After basic quality cuts & 554 & 402 \\
  & \textit{\scriptsize \hspace{1em} $\rightarrow$ Require good quality spectral fit} & & \\
4 & After spectral fit quality check & 479 & 372 \\
  & \textit{\scriptsize \hspace{1em} $\rightarrow$ Select redshift range for UV-optical coverage\textsuperscript{b}} & & \\
5 & After redshift-coverage selection & 308 & 243 \\
  & \textit{\scriptsize \hspace{1em} $\rightarrow$ Final manual inspection and cleaning\textsuperscript{c}} & & \\
\noalign{\smallskip}
\hline
\noalign{\smallskip}
6 & \textbf{Final sample for analysis} & \textbf{276} & \textbf{224} \\
\noalign{\smallskip}
\hline
\end{tabular}
\begin{minipage}{0.95\textwidth} 
\footnotesize
\textit{Note.}---The table summarizes the step-by-step filtering process applied to obtain our final analysis sample. The number of unique sources at each step is determined by cross-matching coordinates within a 2-arcsecond radius.
\textsuperscript{a} The basic quality cuts include three criteria applied simultaneously: (1) continuum signal-to-noise ratio S/N $> 2.5$; (2) host galaxy contribution fraction $< 80\%$; and (3) broad line FWHM $< 12,000$ km s$^{-1}$.
\textsuperscript{b} Redshift ranges were selected based on the typical wavelength coverage of each survey to ensure both 3000\,\AA\ and 5100\,\AA\ fall within the spectrum: SDSS ($0.25 < z < 0.8$), DESI ($0.2 < z < 0.9$), and LAMOST ($0.25 < z < 0.76$).
\textsuperscript{c} Manual screening was performed to remove spectra with poor continuum fits or those where, due to variations in SDSS spectral coverage, the required continuum windows were not actually observed despite the source being within the nominal redshift range.
\end{minipage}
\end{table}



\begin{table*}
\centering
\tiny
\caption{Sample properties.\label{tab:sample_properties}}
\setlength{\tabcolsep}{3.5pt}
\begin{tabular}{l cc ccc cc ccc}
\hline\hline
Name & R.A. & Dec. & Redshift & $\log L_{3000}$ & $\log L_{5100}$ & $\log M_{\rm BH}$ & $\log \lambda_{\rm Edd}$ & 
FWHM(\hb) & Flux(\hb) & S/N (\hb) \\ 
 & [deg] & [deg] & & [erg s$^{-1}$] & [erg s$^{-1}$] & [$M_\odot$] &  & [km s$^{-1}$] & [$10^{-17}$ erg s$^{-1}$ cm$^{-2}$] &  \\
\hline
\noalign{\smallskip}
J081237.89+392810.0 & 123.158 & 39.469 & 0.4531 & 44.05 & 44.11 & 8.50 $\pm$ 0.11 & -1.54 & 6215 $\pm$ 153 & 250.7 $\pm$ 18.2 & 13.8 \\
\noalign{\smallskip}
J142023.53+432921.0 & 215.098 & 43.489 & 0.8256 & 44.82 & 44.88 & 8.91 $\pm$ 0.09 & -1.16 & 5940 $\pm$ 102 & 142.1 $\pm$ 9.8 & 14.5 \\
\noalign{\smallskip}
J091552.17+023450.4 & 138.967 & 2.581 & 0.1220 & 42.98 & 43.15 & 7.95 $\pm$ 0.18 & -1.93 & 3520 $\pm$ 110 & 815.2 $\pm$ 45.1 & 18.1 \\
\hline
\end{tabular}
\begin{minipage}{0.9\textwidth}
\footnotesize
\textit{Note.} The full table is available in machine-readable form in \url{https://doi.org/10.57760/sciencedb.j00167.00048}.
\end{minipage}
\end{table*}



\section{Results}
\label{sect:results}

\begin{figure}
   \centering
   \includegraphics[width=\textwidth]{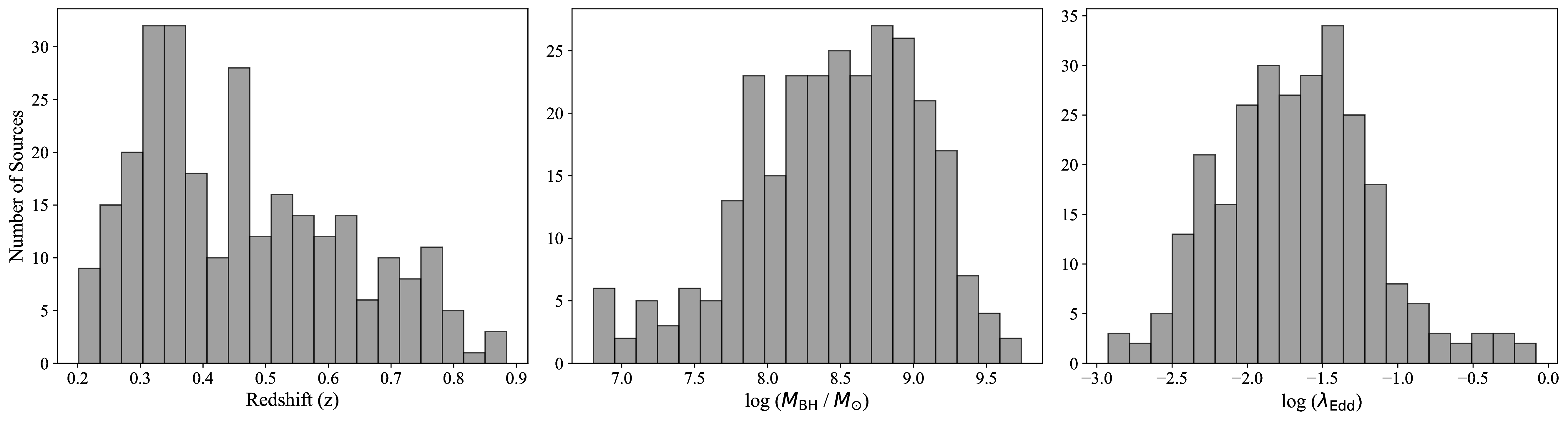}
   \caption{Distributions of redshift($z$, left panel), black hole mass ($M_{\rm BH}$, middle panel), and bolometric Eddington ratio ($\lambda_{\rm Edd}$, right panel) for our sample of 276 sources.}
   \label{fig:sample_dist}
\end{figure}

Figure~\ref{fig:sample_dist} presents the distributions of redshift, black hole mass, and bolometric Eddington ratio for our final sample. The left panel shows that the sources span a redshift range from $z = 0.20$ to $0.88$, with a median redshift of $z \sim 0.44$. The middle panel displays the distribution of our best-estimate black hole masses, which has a median value of $10^{8.50} M_\odot$. The right panel shows the distribution of the bolometric Eddington ratio, which roughly ranges from 0.001 to 0.1 with a  median value of $\log \lambda_{\rm Edd} \sim -1.66$ (i.e., $\lambda_{\rm Edd} \sim 2.2\%$).


\begin{figure}
   \centering
   \includegraphics[width=0.8\textwidth]{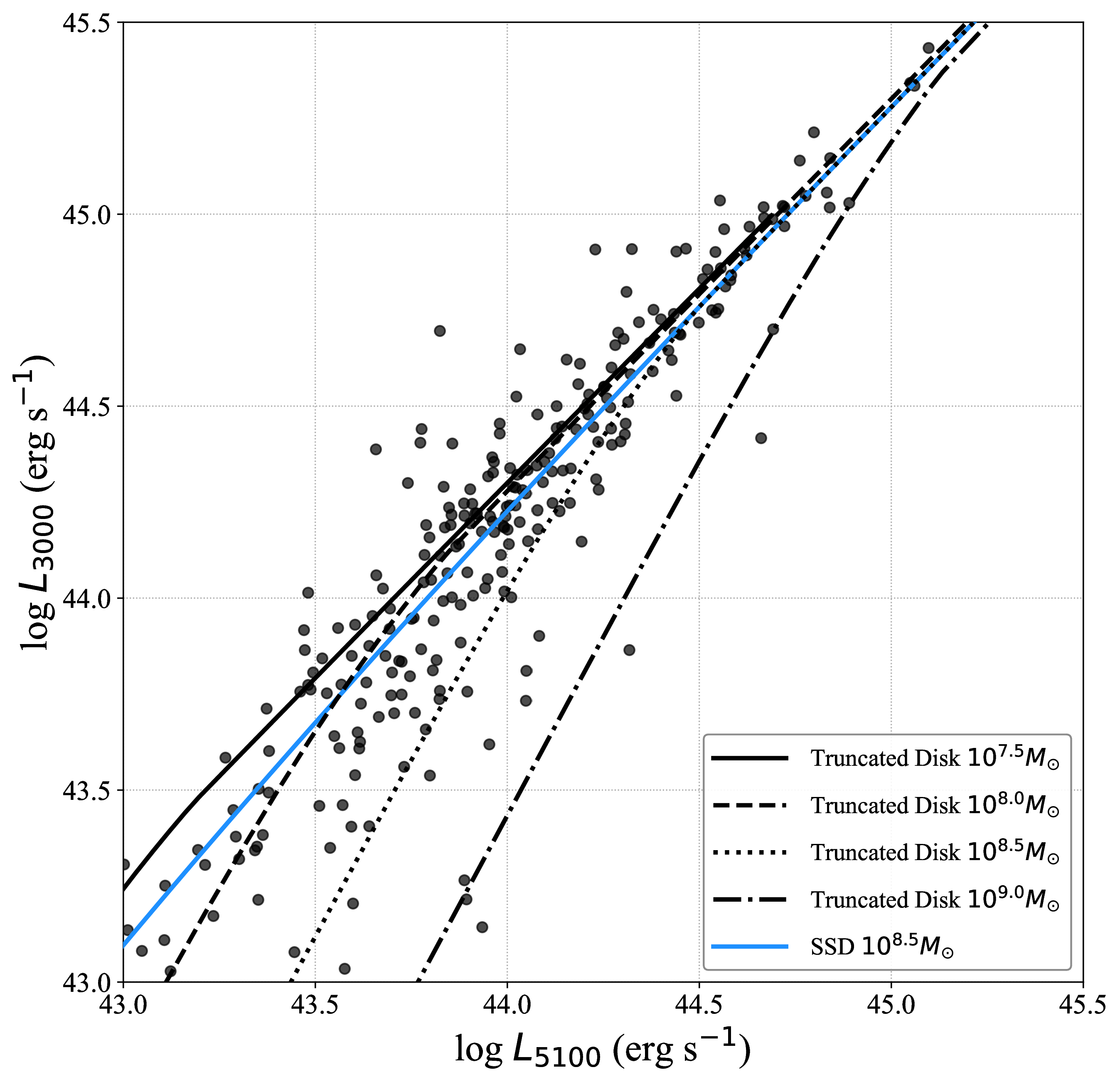}
   
   \caption{The correlation between UV ($L_{3000}$) and optical ($L_{5100}$) continuum luminosity. The gray circles represent the 276 spectra in our sample. The black curves show theoretical predictions from the truncated disk model for four typical black hole masses ($\log M_{\rm BH} = 7.5, 8, 8.5, 9$). The blue curve represents the prediction from a SSD model for a fixed mass of $10^{8.5} M_\odot$.}
   \label{fig:l3000_l5100}
\end{figure}

We present the correlation between the UV and optical continuum luminosities at 3000\,\AA\ ($L_{3000}$) and 5100\,\AA\ ($L_{5100}$) in Figure~\ref{fig:l3000_l5100}. For comparison, we also plotted the prediction from the standard accretion disk (blue solid line with the median value of the SMBH mass in our sample). It can be found that the sources with higher luminosities (e.g.,$L_{3000}>10^{44.5}$erg/s) are roughly consistent with the model prediction. However, most of the sources with low luminosities (e.g.,$L_{3000}<10^{44}$erg/s) deviate from the standard disk model. These low-luminosity sources have the lower 3000\AA\ luminosity at a given 5100\AA\ luminosity. 

\begin{figure}
   \centering
   \includegraphics[width=0.9\textwidth]{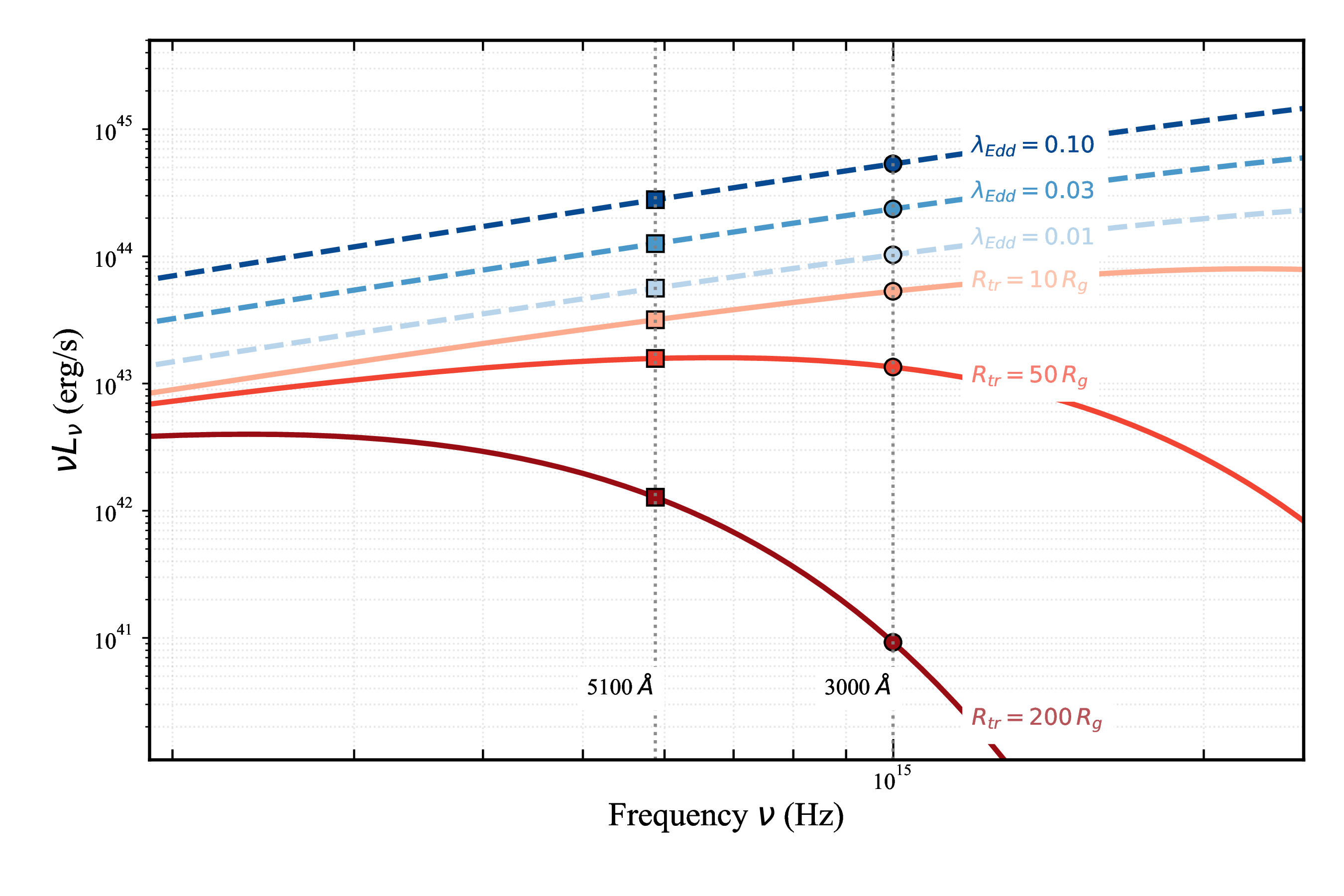}
   \caption{Comparison of theoretical SEDs from a truncated disk model for a $10^8 \, M_{\odot}$ black hole, illustrating the distinct impacts of the inner radius and the accretion rate. The {solid lines} show models with a fixed Eddington ratio (log $\lambda_{\rm Edd} \approx -2.3$) but an increasing inner radius $R_{\rm tr}$. The {dashed lines} show models with a fixed inner radius at the ISCO ($6\,R_{\rm g}$) but a decreasing Eddington ratio.}
   \label{fig:sed_evolution_combined}
\end{figure}

The Shakura-Sunyaev disk (SSD) predicts a self-similar evolution of the disk spectral energy distribution (SED), resulting in a nearly constant optical-to-UV spectral index (see blue dashed lines in Figure \ref{fig:sed_evolution_combined}), which is inconsistent with the observational result at lower luminosities. It is well known that the standard disk will transit to ADAF when the accretion rate is lower than $\sim$1\% Eddington accretion rate, even though the physical mechanism is not fully understood. In this case, the UV emission will quickly become fainter with the increase of truncation radius ($R_{\rm tr}$). To tentatively describe the spectrum from a truncated disk, we employ a color-corrected outer accretion disk model. The radiation of the inner ADAF is neglected, which is normally very faint and mainly radiates at the X-ray waveband. In this work, the truncation radius is described as
\begin{equation}
    R_{\rm tr} = R_{\rm ISCO} \times 
    \begin{cases}
        (\lambda_{\rm Edd} / \lambda_{\rm crit})^{-p} & \text{if } \lambda_{\rm Edd} < \lambda_{\rm crit} \\
        1 & \text{if } \lambda_{\rm Edd} \geq \lambda_{\rm crit},
    \end{cases}
    \label{eq:Rin}
\end{equation}
where $R_{\rm ISCO}$ corresponds to innermost stable circular orbit, $p$ is a power-law index, and $\lambda_{\rm crit}$ is the critical Eddington ratio for the accretion-mode transition. We adopted a canonical value of $\lambda_{\rm crit} = 0.02$ and a physically motivated index of $p=2$ \citep[e.g.,][]{2018MNRAS.480.3898N}.  We noted that the shape of the theoretical tracks in the $L_{3000}$–$L_{5100}$ plane is not sensitive to the precise value of $p$, and our results remain largely unchanged for different choices (e.g., $p=1$ or $p=3$). The SED can be derived by integrating the blackbody emission from each annulus of the disk, where we also considered a color correction factor $f_{\rm col}$ to account for atmospheric effects \citep{2002ApJ...572...79C,2022ApJ...939L...2Z}. In Figure \ref{fig:sed_evolution_combined}, we presented an example of the spectrum from the truncated disk with $\lambda_{\rm Edd} = 5\times 10^{-3}$ and truncation radius of 10, 50 and 200 $R_{\rm g}$ ($R_{\rm g}=GM_{\rm BH}/c^2$ is gravitational radius, $G$ is gravitational constant and $c$ is light speed). It can be seen that the UV emission declines much faster than the optical waveband. The predicted spectra from the truncated disk with several typical black hole masses are presented in Figure \ref{fig:l3000_l5100}. The $L_{3000}-L_{5100}$ relation becomes steeper at lower luminosities, which is roughly consistent with the observational results of CLAGNs.

\section{Discussion}
\label{sect:discussion}

{ The $L_{3000}$-$L_{5100}$ correlation of most of CLAGNs deviates from the prediction of standard accretion disk and is statistically consistent with the truncated disk model. The Eddington ratios of most CLAGNs are less than several percent, which are similar to the low-hard state X-ray binaries and the typical low-luminosity AGNs, where the inner standard accretion disk is believed to truncated into hot ADAF \citep{1994ApJ...428L..13N}. The SED modelling with high-resolution observational data of Mrk 1018 indeed supports this scenario \citep{2018MNRAS.480.3898N,2021MNRAS.506.4188L}. It should be noted that the data quality is not so good for these low-luminosity AGNs based on the ground-based SDSS/DESI/LAMOST observations. Furthermore, the truncated disk model can not be well constrained with the narrow-band observations at 3000$\rm \AA$ and 5100$\rm \AA$. The high-sensitivity and high-resolution UV band observation will provide much better constraints on theoretical models. In the time-domain era, more spectroscopic observations with photometric monitoring for more CLAGNs will also shed light on this issue, where the better data on high state and dim state can be used to model the disk evolution.    
}

\subsection{The role of ionizing photons in driving the changing look}
The defining characteristic of a changing-state AGN is the appearance or disappearance of its broad emission lines. These lines are produced by gas in the BLR that is photoionized by the central engine's continuum emission. The strength of the emission lines is critically dependent on the flux of high-energy, ionizing photons, which primarily originate from the innermost, hottest regions of the accretion disk. Our results show that as a CLAGN transitions to a dim state, the disk truncation leads to a catastrophic drop in the UV continuum. This mechanism provides a direct causal link between the change in accretion mode and the observed ``changing look", consistent with the findings of many recent CLAGN surveys \citep[e.g.,][]{2025ApJS..278...28G, 2025ApJ...986..160D}.  \citet{2023ApJ...950..106W} explored the evolution of broad lines based on the possible evolution of ionization spectrum and CLOUDY modelling, where the anti-correlation between Balmer decrement and Eddington ratio can be well explained by the evolution of the accretion disk. It should be noted that we didn't model the spectrum of each CLAGN in our sample, which is mainly caused by the narrow-band spectra and possible parameter degeneracy. The broadband spectra with both optical-UV and X-ray observations will shed further light on the modelling \citep[e.g.,][]{2018MNRAS.480.3898N}.

\subsection{The Physics of Disk Transitions and the Timescale Problem}
A long-standing puzzle in AGN variability is that the observed timescales for dramatic changes, typically years \citep[e.g.,][]{2020MNRAS.491.4925G}, are often much shorter than the viscous timescale predicted for the bulk of the accretion disk, which can be thousands of years or longer \citep[e.g.,][]{2018NatAs...2..102L,2018MNRAS.480.3898N}. { The viscous crisis might be partly solved when considering the large-scale magnetic fields with magnetic pressure supported thick disk model \citep{2019MNRAS.483L..17D} or the magnetic outflows driven disk-outflow model} \citep{2021ApJ...916...61F}. 
Our sample of CLAGNs has Eddington ratios clustering around the critical value of $\lambda_{\rm Edd} \sim 2.2\%$ (see Fig.~\ref{fig:sample_dist}). 
At this threshold, the accretion disk is expected to be thermally unstable. The disk is neither a simple, stable SSD nor a stable ADAF, but exists in a regime where it is susceptible to rapid state transitions. A relatively small perturbation could trigger a rapid collapse of a hot inner flow into a thin disk \citep[a turn-on event; ][]{2025ApJ...988..207L} or the evaporation of an inner thin disk into a hot flow \citep[a turn-off event; e.g.,][]{2010ApJ...724..855C} on a much shorter, thermal timescale \citep[][]{1994A&A...288..175M}, with some transitions observed to occur in as little as a few months \citep{2022ApJ...939L..16Z}. { It could be consistent with the  thermal and heating/cooling front timescales correlated to the changes in the innermost regions of the accretion disk} \citep[e.g.,][]{2018MNRAS.480.4468R,2018ApJ...864...27S}. The fact that CLAGNs are preferentially found at this critical Eddington ratio strongly supports the idea that we are witnessing these rapid, instability-driven state transitions in action. This idea is further strengthened by the growing population of repeating or recurrent CLAGNs, which have been observed to transition between types multiple times \citep[e.g.,][]{2025A&A...693A.173L, 2025arXiv251018445D, 2025ApJ...981..129W, 2025A&A...698A.135G}, suggesting that the changing-look state is a recurring phase of AGN activity. {Such a long timescale might be shortened by partial instability in a narrow unstable zone \citep{2020A&A...641A.167S} or large-scale magnetic fields driven wind \citep{2021ApJ...910...97P}. }

\subsection{The Importance of UV Monitoring}
Our results highlight the crucial role of the UV band in understanding the CLAGN phenomenon. As shown in Figure~\ref{fig:sed_evolution_combined}, a recession of the inner disk from a few to several tens of gravitational radii can cause the UV flux (e.g., at 3000\,\AA) to drop by orders of magnitude, while the optical flux (e.g., at 5100\,\AA) may change by a factor of only a few \citep[e.g.,][]{2016A&A...593L...9H,2018MNRAS.480.3898N}. Therefore, optical monitoring alone may miss the most dramatic part of the accretion flow transformation or underestimate its magnitude. It should be noted that most CLAGNs in faint state should have lower 3000\AA\ luminosity (stay in bottom-left region in Figure~\ref{fig:l3000_l5100}), which are not included because their UV emission cannot be well constrained in the low state. The high-sensitivity UV observations will put further constraints on accretion flow in the low-luminosity regime in the future. In addition, the high-cadence monitoring campaigns in the UV band will be essential for capturing the onset of these state transitions, precisely measuring their timescales, and providing the most stringent tests of accretion disk instability models \citep[e.g.,][]{2023A&A...672A..19S}.

\section{Summary}
\label{sect:conclusion}

In this work, we investigated the UV-optical continuum luminosity correlation of a sample of CLAGNs using multi-epoch spectra from SDSS, DESI, and LAMOST. Our analysis leads to the following main conclusions:

\begin{enumerate}
    \item We discovered a systematic steepening in the $L_{3000}$--$L_{5100}$ correlation for CLAGNs at low luminosities, indicating that the AGN SED becomes significantly redder as the source dims. This behavior is inconsistent with the self-similar evolution predicted by the standard accretion disk model.
    \item We demonstrated that this observed trend is in good agreement with the predictions of a truncated accretion disk model, where the inner disk recedes as the accretion rate falls below a critical value.
    \item Our results provide strong statistical evidence that the CLAGN phenomenon is driven by a global state transition in the accretion flow. The disappearance of the inner disk quenches the supply of ionizing photons, leading directly to the fading of the broad-line region and the observed spectral type change.
\end{enumerate}

\begin{acknowledgements}
We acknowledge the anonymous referee for constructive comments that improve the paper. We thank Wei-jian Guo and Qian Dong for providing their changing-look AGN catalogs, which were essential for compiling our parent sample. The work is supported by National Natural Science Foundation of China (grants 12533005, 12233007).This work has made use of data from the Sloan Digital Sky Survey (SDSS). Funding for the SDSS IV has been provided by the Alfred P. Sloan Foundation, the U.S. Department of Energy Office of Science, and the Participating Institutions. SDSS-IV acknowledges support and resources from the Center for High-Performance Computing at the University of Utah. The SDSS website is www.sdss.org.This paper makes use of data from the Dark Energy Spectroscopic Instrument (DESI), a program of the NSF's NOIRLab. DESI is a primary funding from the U.S. Department of Energy, Office of Science. NOIRLab is managed by the Association of Universities for Research in Astronomy (AURA) under a cooperative agreement with the National Science Foundation. We acknowledge Guoshoujing Telescope (the Large Sky Area Multi-Object Fiber Spectroscopic Telescope, LAMOST), which is a National Major Scientific Project built by the Chinese Academy of Sciences. Funding for the project has been provided by the National Development and Reform Commission. LAMOST is operated and managed by the National Astronomical Observatories, Chinese Academy of Sciences.

\textbf{Software:} Astropy \citep{2022ApJ...935..167A} and \texttt{PyQSOFit} \citep{2018ascl.soft09008G, 2019ApJS..241...34S}.
\end{acknowledgements}

\bibliographystyle{raa}
\bibliography{bibtex}

\end{document}